\documentclass{LMCS}

\usepackage{enumerate,proof,hyperref}
\def\eqalign#1{\null\,\vcenter{\openup\jot\mathsurround=0 pt
  \ialign{\strut\hfil$\displaystyle{##}$&$\displaystyle{{}##}$\hfil
      \crcr#1\crcr}}\,}

\def\doi{2 (4:5) 2006}
\lmcsheading%
{\doi}
{1--17}
{}
{}
{Dec.~\phantom{0}9, 2005}
{Oct.~18, 2006}
{}

\begin{document}

\title[Solution of a Problem of Barendregt]{Solution of a Problem of
Barendregt on Sensible $\lambda$-Theories}

\author[B.~Intrigila]{Benedetto Intrigila\rsuper{a}}   
\address{{\lsuper a}Universit\`{a} degli Studi di Roma "Tor Vergata" \\
          Rome, Italy} 
\email{intrigil@mat.uniroma2.it}  

\author[R.~Statman]{Richard Statman\rsuper{b}} 
\address{{\lsuper b}Carnegie-Mellon University \\ Pittsburgh, PA,
USA}  
\email{rs31@andrew.cmu.edu}  



\begin{abstract}
  $\mathcal{H}$ is the theory extending $\beta$-conversion by
identifying all closed unsolvables. $\mathcal{H}\mathbf{\omega}$ is
the closure of this theory under the $\omega$-rule (and
$\beta$-conversion). A long-standing conjecture of H. Barendregt
states that the provable equations of $\mathcal{H}\mathbf{\omega}$
form a $\mathbf{\Pi_{1}^{1}}$-complete set. Here we prove that
conjecture.

\end{abstract}

\maketitle

\section{Introduction}\label{S:one}

 There is a strong need to make theorem provers such as COQ or
ISABELL/HOL more and more powerful (see e.g. \cite{COQ}, \cite{ISA},
\cite{HOL}). In particular it seems very hard to automatically set up
inductive arguments to get universal conclusions. In this sense, the
use of some (constructive) kind of $\omega$-rule is very appealing
since one could get a universal conclusion from, say, a finite number
of cases. Typically, this happens when for every property $P$ of
interest, there exists a computable upper bound $k$ such that if every
ground term of complexity less than $k$ satisfies $P$ then $\forall
x. P(x)$ holds, so that a universal conclusion can be obtained e.g.
by a systematic search on a finite set of cases.

Therefore, it is important to precisely assess the logical power of
the $\omega$-rule in the different computational contexts. Here we
consider the $\omega$-rule in the $\lambda\beta$-calculus. We have
already considered constructive forms of such rule in \cite{IS2002},
obtaining recursively enumerable $\lambda$-theories which are closed
under the $\omega$-rule. Moreover, in \cite{IS2004}, we have
considered the more important problem of the $\omega$-rule added to
the {\em pure} $\lambda\beta$-calculus. We have shown that the
resulting theory is {\em not} recursively enumerable, by giving a
many-one reduction of the set of true $\mathbf{\Pi_{2}^{0}}$ sentences
to the set of consequences of the lambda calculus with the
$\omega$-rule. This solved in the affirmative a well known problem of
H. Barendregt \cite{Ba84, FlaggMyhill}. More recently we have obtained
the result (\cite{IS2005}) that such theory is not even {\em
arithmetical}.

Here we consider the problem of determining the computational power
of the $\omega$-rule added to the theory $\mathcal{H}$.
$\mathcal{H}$ is the theory obtained extending $\lambda\beta$ by
identifying all closed unsolvable terms. We prove that the resulting
theory $\mathcal{H}\mathbf{\omega}$ is
$\mathbf{\Pi_{1}^{1}}$-complete, which solves another long-standing
conjecture of H. Barendregt (see \cite{Ba84} Conjecture 17.4.15).

\section{The System $\mathcal{H}\mathbf{\omega}$}\label{two}

Notation will be standard and we refer to \cite{Ba84}, for
terminology and results on $\lambda$-calculus. In particular:
\begin{enumerate}[$\bullet$]
\item $\equiv$ denotes syntactical identity;
\item by the notation $[X/x]Y$ we mean the replacement of term $X$
for the variable $x$ inside $Y$, with the usual proviso that no free
variable $y$ of $X$, with $y \not\equiv x$, becomes bounded after
the substitution (see \cite{Ba84} 2.1.11-2.1.14);
\item $\longrightarrow_{\beta}$, $\longrightarrow_{\eta}$ and
$\longrightarrow_{\beta\eta}$
 denote  $\beta$-, $\eta$- and, respectively, $\beta\eta$-reduction and
$\longrightarrow^{*}_{\beta}$ $\longrightarrow^{*}_{\eta}$ and
$\longrightarrow^{*}_{\beta\eta}$ their respective reflexive and
transitive closures;
\item $=_{\beta}$ and $=_{\beta\eta}$ denote  $\beta$- and, respectively, ${\beta\eta}$-conversion;
\item combinators (i.e. closed $\lambda$-terms)  such e.g. $\mathbf{I}$ have the usual meaning;
\item $\underline{k}$ denotes the $k$-th Church numeral.\end{enumerate}
$\lambda$-terms are denoted by capital letters: in particular we
adopt the convention that $M,N,P,$ $Q,\ldots$ are {\em closed} terms
and $U,V,X,Y,W,Z$ are possibly {\em open} terms.

The notion of {\em $\lambda$-theory} has the usual meaning of
\cite{Ba84} Ch.4, that is a consistent set of equations between
closed terms, which is closed under the axioms and the rules of
$\lambda\beta$-calculus. We now briefly recall the
$\lambda$-theories we are concerned with.

By $\lambda \beta$ we denote pure $\beta$-convertibility (see
\cite{Ba84}). $\mathcal{H}$ is the $\lambda$-theory extending
$\lambda\beta$ by identifying all closed unsolvable terms, see
\cite{Ba84} Definition 4.1.6. We recall that this $\lambda$-theory
can be formulated by adding to $\lambda\beta$ all equations of the
form $M = \mathbf{\Omega}$, where $M$ is a closed unsolvable term,
the combinator $\mathbf{\Omega}$ is defined as
$\mathbf{\omega}\mathbf{\omega}$ and $\mathbf{\omega}$ is $\lambda
x.xx$. Moreover, we recall also that $\mathcal{H}$ is generated by
the notion of reduction $\mathbf{\beta}\mathbf{\Omega}$, see
\cite{Ba84} Lemma 16.1.2. The notion of reduction
$\mathbf{\beta\Omega}$ is defined by adding to the $\beta$-reduction
rule, the (non constructive)
reduction rule:
\[M \longrightarrow \mathbf{\Omega}\quad\hbox{if $M$ is unsolvable and $M
\not\equiv \mathbf{\Omega}$}\]
 see \cite{Ba84} Section 15.2.

$\mathcal{H}\mathbf{\omega}$ is the $\lambda$-theory obtained by
adding the so called {\em $\omega$-rule} to $\mathcal{H}$, see
\cite{Ba84} Definition 4.1.10 and Paragraph 4.2.

We formulate $\mathcal{H}\mathbf{\omega}$ differently. As the reader
will see, we want a formulation of the theory such that only
equalities between {\em closed} terms can be proven.

\begin{defi}\label{Ho_def}{\em Equality in
  $\mathcal{H}\mathbf{\omega}$ } (denoted by
  $=_{\omega}$) is defined by the following axioms and rules:
\begin{enumerate}[(1)]
\item {\em Identity Axioms}\label{identity}:
                \[M =_{\omega} M\]
\item {\em Weak $\mathbf{\beta}\mathbf{\Omega}$-Conversion Axioms}\label{wbo}:
\[\eqalign{
  (\lambda x.U) N
&=_{\omega} [N/x]U\cr
  [N/x]U
&=_{\omega} [N/x]U\cr
  M
&=_{\omega}\mathbf{\Omega}\cr
  \mathbf{\Omega}
&=_{\omega} M\cr
}
\qquad
\eqalign{
&\hbox{(with $(\lambda x.U) N$ closed)}\cr
&\hbox{(with $(\lambda x.U) N$ closed)}\cr
&\hbox{(with $M$ closed and unsolvable)}\cr
&\hbox{(with $M$ closed and unsolvable)}\cr
}
\]

\item {\em Leibnitz Rule: Substitute Equals for Equals}\label{leibnitz}:

\[\infer{[N/z]X =_{\omega}[N/z]Y}
        {[M/z] X =_{\omega} [M/z]Y\qquad M =_{\omega} N}
\]
  where terms $X$ and $Y$ have possibly $z$ as free variable, and no
  other free variable.
\item {\em The $\omega$-Rule}\label{omega}:
\[\infer{P=_{\omega}Q}
        {\forall M,\ M\hbox{\ closed,\ }PM =_{\omega} QM}
\]
\end{enumerate}
\end{defi}

We call $\mathcal{H}\mathbf{\omega}$ the $\lambda$-theory specified
above. In the next Section we prove that this formulation
gives rise to the same theory of \cite{Ba84}.

\section{Derived Rules}\label{three}
Now we prove that some rules are derived rules in
$\mathcal{H}\mathbf{\omega}$.

\begin{prop}(Symmetry)\label{simmetry}
For every $M$ and $N$, if $M =_{\omega}
N$ then $N =_{\omega} M$.
\end{prop}

\proof Axioms and rules of $\mathcal{H}\mathbf{\omega}$ are completely
symmetric, so a proof of $M =_{\omega} N$ can be converted into one of
$N =_{\omega} M$ by reversing sides.\qed

\begin{prop}(Transitivity)\label{transitivity}
 For every $M$, $N$ and $P$, if $M =_{\omega}
N$ and $N =_{\omega} P$ then $M =_{\omega} P$.
\end{prop}

\proof
Assume $M =_{\omega}N$. So, $N =_{\omega}M$. Then let $X$ be $z$ and
$Y$ be $M$. We have that $[N/z] X =_{\omega} [N/z]Y$ and
\[\infer{P =_{\omega} M}
        {[N/z] X =_{\omega} [N/z]Y\qquad N =_{\omega} P}
\]
by one application of the Leibnitz Rule.\qed

By $\mathbf{\beta\Omega}$-convertibility we mean the convertibility
relation generated by the $\mathbf{\beta\Omega}$-reduction mentioned
above (see \cite{Ba84} Paragraph 15.2). By a context $Z[\ ]$ we mean
a term with {\em holes} in the sense of \cite{Ba84} Definition
2.1.18.

\begin{prop}
For closed $M$ and $N$, if $M$ $\mathbf{\beta\Omega}$-converts to
$N$ then $M =_{\omega} N$.
\end{prop}

\proof\hfill
\begin{enumerate}[(1)]
\item
Let a context $Z[\ ]$  and terms $\lambda x.U$, $V$ be given. Let
$z_1 ... z_t$ be an enumeration of all free variables occurring in
$Z[\ ]$, $\lambda x.U$ and $V$.

By induction on the complexity of $Z[\ ]$ one can prove that for all
closed $P_1\dots P_t$ we have:
\[[P_1 / z_1, ... ,P_t / z_t]Z[(\lambda x.U) V]
=_{\omega} [P_1 / z_1, ... ,P_t / z_t] Z[[V/x]U]\ .
\]
Thus for $\lambda$-closures
                $\lambda z_1 ... z_t . Z[(\lambda x.U) V]$ and
                $\lambda z_1 ... z_t . Z[[V/x]U]$,
                 we have that for all closed  $P_1 ... P_t$:
\[(\lambda z_1 ... z_t .
Z[(\lambda x.U) V]) P_1 ... P_t =_{\omega}  (\lambda z_1 ... z_t .
Z[[V/x]U]) P_1 ... P_t\ ,
\]
 so $\lambda z_1 ... z_t . Z[(\lambda x.U) V] =_{\omega} \lambda z_1
... z_t . Z[[V/x]U]$ by $t$ applications of the $\omega$-rule.

\item Let a context $Z[\ ]$  and an unsolvable term $V$ be given. Let $z_1
... z_t$ be an enumeration of all free variables occurring in $Z[\
]$ and $V$.

By induction on the complexity of $Z[\ ]$ one can prove that for all
closed $P_1 ... P_t$ we have $[P_1 / z_1, ... ,P_t / z_t]Z[V]
=_{\omega} [P_1 / z_1, ... ,P_t / z_t] Z[\mathbf{\Omega}]$. Thus for
$\lambda$-closures: $\lambda z_1 ... z_t . Z[ V]$ and $\lambda z_1
... z_t . Z[\mathbf{\Omega}]$ , we have that for all closed $P_1
... P_t$:
\[(\lambda z_1 ... z_t .  Z[V ]) P_1 ... P_t =_{\omega}
(\lambda z_1 ... z_t .  Z[\mathbf{\Omega}]) P_1 ... P_t
\]
 so $\lambda z_1 ... z_t .  Z[ V ] =_{\omega} \lambda z_1 ... z_t
. Z[\mathbf{\Omega}]$ by $t$ applications of the $\omega$-rule.\qed
\end{enumerate}

\begin{prop}
 If $X$  $\mathbf{\beta\Omega}$-converts to $Y$
 then the $\lambda$-closures of $X$ and $Y$ are
    provably equal in $\mathcal{H}\mathbf{\omega}$.
\end{prop}

\proof The proposition follows directly from the previous one.\qed

By the previous results it follows that $\mathcal{H}\mathbf{\omega}$
is exactly the same theory defined in \cite{Ba84}.

We observe also the following:

\begin{prop} The so-called $\eta$-conversion (that is $(\lambda x.Mx)
= M$) obviously holds in $\mathcal{H}\mathbf{\omega}$, for any closed
term $M$.  (With respect to the usual formulation of the
$\eta$-conversion, observe that since $M$ is closed there is no need
to require $x$ fresh in $M$).\qed
\end{prop}

\section{Weak $\mathbf{\beta}\mathbf{\Omega}$-Reduction}\label{three.5}

We call {\em weak $\mathbf{\beta}\mathbf{\Omega}$-conversion} the
smallest congruence relation containing the equations of Weak
$\mathbf{\beta}\mathbf{\Omega}$-Conversion Axioms above (see
Definition \ref{Ho_def}.\ref{wbo}). Observe that this relation also
includes the Identity Axioms. We write $M \sim_{w\beta\Omega}
N$ to denote the weak $\beta\mathbf{\Omega}$-conversion relation.
Moreover such equations can be oriented, giving rise to the
following contraction rules:

\noindent {\em weak $\beta$-contraction rule}
\[(\lambda x.M) N \longrightarrow_{w\beta} [N/x]M
  \qquad\hbox{(with $(\lambda x.M) N$ closed)}\]

\noindent {\em weak $\mathbf{\Omega}$-contraction rule}
\[M \longrightarrow_{\Omega} \mathbf{\Omega}
  \qquad\hbox{(with $M$ closed and
unsolvable and $M \not\equiv \mathbf{\Omega}$)}\]

We call {\em weak $\beta\mathbf{\Omega}$-reduction} the reduction
relation generated by the two rules, after closure under contexts
(see \cite{Ba84} 3.1). It is easy to see that the weak
$\mathbf{\beta}\mathbf{\Omega}$-conversion is the convertibility
relation generated by the weak $\beta\mathbf{\Omega}$-reduction. We
shall call the two contraction rules above also {\em weak
$\beta$-reduction rule} and, respectively {\em weak
$\mathbf{\Omega}$-reduction rule}. This terminology includes the
case that such rules are applied inside a context.\\
We write:
\[\longrightarrow_{w\beta\Omega}\quad\hbox{and}\quad\longrightarrow^{*}_{w\beta\Omega}\]
to denote weak $\beta\mathbf{\Omega}$-reduction and, respectively, its
reflexive and transitive closure.

We recall a result needed in the following.

\begin{prop}
Weak $\beta\mathbf{\Omega}$-reduction is Church-Rosser.
\end{prop}

\proof
Weak $\beta$- and, respectively, weak $\mathbf{\Omega}$-reductions
are both Church-Rosser and commute. Now, use the Hindley-Rosen Lemma
(see \cite{Ba84}, 3.3.5).\qed

In the sequel, we shall need the following notions on reductions. We
define the notions of {\em trace} and {\em extended trace (etrace)}
as follows. Given the reduction $F \longrightarrow_{\beta}^{*} G$
(or the reduction $F \longrightarrow_{w\beta\Omega}^{*} G$) and the
closed subterm $M$ of $F$, the {\em traces} of $M$ in the terms of
the reduction are simply the copies of $M$ until each is either
 deleted by a contraction of a redex with a dummy lambda, replaced by
 $\mathbf{\Omega}$ by an $\mathbf{\Omega}$-reduction (possibly of
 a superterm $M'$ of $M$) or altered by a reduction internal to $M$
 or a reduction with $M$ at
the head (when $M$ begins with lambda or when the reduction is an
$\mathbf{\Omega}$-reduction). The notion of {\em etrace} is the same
except that we allow internal reductions, so that a copy of $M$
altered by an internal reduction continues to be an etrace.

\section{Normal Form for $\mathcal{H}\mathbf{\omega}$ Proofs}\label{four}

 As usual proofs in $\mathcal{H}\mathbf{\omega}$ can be thought
of as (possibly infinite) well-founded trees. We distinguish between
two cases.
\begin{enumerate}[$\bullet$]
\item The proof ends with an application of the $\omega$-rule.
\item Otherwise. So, we can consider all nodes of the proof tree
that have no premises of the $\omega$-rule as descendant. Or, in
other terms, there are no occurrences of the $\omega$-rule in the
path from the node to the conclusion of the proof. We call the set
of such nodes the {\em endpiece} of the proof.
\end{enumerate}

Notice that the endpiece of a proof consists of a finite tree of
Leibnitz Rule inferences all of whose leaves are either instances of
the Identity Axioms , instances of the Weak
$\mathbf{\beta}\mathbf{\Omega}$-Conversion Axioms, or direct
conclusions of the $\omega$-rule. The tree reduces to a single node
in case the proof amounts to an instance of the Identity Axioms or
to an instance of the Weak
$\mathbf{\beta}\mathbf{\Omega}$-Conversion Axioms. We shall put this
endpiece into a normal form.

\begin{defi}\label{e-p-nf-def}
An endpiece is {\em
in normal form} iff it is of the form:
\[\infer{M =_{\omega} N}
  {\infer{M =_{\omega} M_t Q_t}
   {\infer{\dots\strut}
    {\infer{M =_{\omega} M_2 Q_2}
     {\infer{M =_{\omega} M_1 Q_1}
      {\infer{M =_{\omega} M_1 Q_1}
       {M =_{\omega} M
      &M \sim_{w\beta\Omega} M_1 P_1}
     &\mkern-16 mu P_1 =_{\omega} Q_1}
    &\mkern-16 mu M_1 Q_1 \sim_{w\beta\Omega} M_2 P_2}
   &\mkern-16 mu P_2 =_{\omega} Q_2}
  &\mkern-16 mu\hbox to 10 pt{\hfill}}
 &\mkern-16 mu M_t Q_t \sim_{w\beta\Omega} N}
\]
where each equality of the form $P_i =_{\omega} Q_i$, for
$1 \leq i \leq t$, is a direct conclusion of the $\omega$-rule.\\
 We allow the {\em degenerate case} $t = 0$ and consider {\em in normal form}:
 \begin{enumerate}[$\bullet$]
\item an instance of
the Identity Axioms;
\item as well as the endpiece:
\[\infer{M =_{\omega} N}
        {M =_{\omega} M\qquad M \sim_{w\beta\Omega} N}
\]
 \end{enumerate}
\end{defi}

\noindent{\bf Remark.} In the previous Definition, observe that the
intuitive motivation of the notion of {\em normal form} is to have a
mean to separate - into the endpiece of a proof - the conclusions of
the $\omega$-rule from the other components of the endpiece itself.
The normal form diagram represents a sequence of applications of the
Leibnitz Rule, and in particular the odd lines are instances of the
transitivity rule, while the even ones are substitutions of the term
$Q_i$ for the equal term $P_i$ in the applicative context $M_i[\ ]$.
Observe that the normal form diagram is {\em not}, strictly
speaking, a proof tree in $\mathcal{H}\mathbf{\omega}$, since
premises of the form $M_i Q_i \sim_{w\beta\Omega} M_{i+1} P_{i+1}$
refer to the $\mathbf{\beta}\mathbf{\Omega}$-conversion relation.
(This explains why we need two {\em degenerate} cases, one when the
whole proof is an instance of Identity Axioms, and the other one
when we have a proof of the
$\mathbf{\beta}\mathbf{\Omega}$-convertibiliy of the terms $M$ and
$N$. Of course, the former could also be considered a particular
case of the latter). It is clear, however, by the result of Section
\ref{three}, that each $\mathbf{\beta}\mathbf{\Omega}$-conversion
can be expanded into a $\mathcal{H}\mathbf{\omega}$ proof tree.

\begin{thm}\label{end-p-nf}
For every proof $\mathcal{T}$ in $\mathcal{H}\mathbf{\omega}$ there
exists a proof $\mathcal{T'}$ of the same conclusion with the
endpiece in normal form.
\end{thm}

We shall use several lemmata.
\begin{lem}\label{sym_trans}
Proofs in normal forms are closed under symmetry and transitivity
rules.
\end{lem}

\proof\hfill
\begin{enumerate}[$\bullet$]
\item (symmetry)

\noindent To see this, observe that a proof in normal form of $ M =_{\omega}
N$ can be reversed into a proof in normal form of $ N =_{\omega} M$.
Indeed, each direct conclusion of the $\omega$-rule $P_i =_{\omega}
Q_i$ can be reversed in a direct $\omega$-rule conclusion of $Q_i
=_{\omega} P_i$, since for every premise $P_i R =_{\omega} Q_i R$
there is (by Proposition \ref{simmetry}) a premise $Q_i R =_{\omega}
P_i R$. Moreover the sequences $Q_t, Q_{t-1} \ldots Q_1$ and $P_t,
P_{t-1} \ldots P_1$ take the place of $P_1, P_2 \ldots P_t$ and,
respectively, of $Q_1, Q_{2} \ldots Q_t$.

\item (transitivity)

\noindent Given proofs in normal form of $M =_{\omega} N$ and of $N =_{\omega}
P$, to obtain a proof in normal form of $M =_{\omega} P$ contract
the last row of $M =_{\omega} N$:
\[\infer{M =_{\omega} N}
        {M =_{\omega} M_t Q_t\qquad M_t Q_t \sim_{w\beta\Omega} N}
\]
and the first row of $N =_{\omega} P$:
\[\infer{N =_{\omega} M'_1 Q'_1}
        {N =_{\omega} N\qquad N \sim_{w\beta\Omega} M'_1 Q'_1}
\]
as follows:
\[\infer{M =_{\omega} M'_1 Q'_1}
        {M =_{\omega} M_t Q_t\qquad M_t Q_t \sim_{w\beta\Omega} M'_1 Q'_1}
\]
and then follows the proof in normal form of $ N =_{\omega} P$
replacing every left-side occurrence of $N$ with $M$.\qed
\end{enumerate}

\begin{lem}\label{context}
If there is a proof in normal form of $ M =_{\omega} N$ then for
every $X$, with a unique free variable $z$, there is a proof in
normal form, with the same length, of $[M/z]X =_{\omega} [N/z]X$.
\end{lem}

\proof
Let the proof of $ M =_{\omega} N$ be of the form:
\[\infer{\dots\strut}
  {\infer{M =_{\omega} M_2 P_2}
   {\infer{M =_{\omega} M_1 Q_1}
    {\infer{M =_{\omega} M_1 P_1}
           {M =_{\omega} M&M \sim_{w\beta\Omega} M_1 P_1}
          &P_1 =_{\omega} Q_1}
         &M_1 Q_1 \sim_{w\beta\Omega} M_2 P_2}}
\]
  To get a proof in normal form of $[M/z]X =_{\omega} [N/z]X$,
  transform it as follows:
\[\infer{\dots\strut}
  {\infer{[M/z]X =_{\omega} (\lambda u.(\lambda z. X)(M_2 u)) P_2}
   {\infer{[M/z]X =_{\omega} (\lambda u.(\lambda z. X)(M_1 u)) Q_1\mkern75 mu}
    {\infer{[M/z]X =_{\omega} (\lambda u.(\lambda z. X)(M_1 u)) P_1\mkern180 mu}
           {[M/z]X =_{\omega} [M/z]X\mkern100 mu&
 \mkern-100 mu[M/z]X \sim_{w\beta\Omega} (\lambda u.(\lambda z. X)(M_1 u)) P_1}
          &\mkern-180 mu P_1 =_{\omega} Q_1}
         &\mkern-75 mu (\lambda u.(\lambda z. X)(M_1 u)) Q_1 \sim_{w\beta\Omega} (\lambda u.(\lambda z. X)(M_2 u)) P_2}}
\]
\qed

\begin{lem}
Proofs with the endpiece in normal form are closed under Axioms and
Rules of $\mathcal{H}\mathbf{\omega}$ .
\end{lem}

\proof
We argue by induction on the complexity of the proof of $M
=_{\omega} N$.
\begin{enumerate}[(1)]
\item If the proof consists of an instance of the identity axiom,
then there is nothing to prove.

\item If the proof consists of an instance of the Weak
 $\beta\mathbf{\Omega}$-Conversion Axioms, then use the fact that the
 sequences $P_1, \dots , P_t$ and $Q_1, \dots , Q_t$ can be empty
 (i.e. $t=0$).

\item If the proof ends with an instance of the $\omega$-rule then
it can be put in normal form as follows:
\[\infer{M =_{\omega} N}
  {\infer{M =_{\omega} \mathbf{I} N}
   {\infer{M =_{\omega} \mathbf{I} M}
    {M =_{\omega} M
   &M \sim_{w\beta\Omega} \mathbf{I} M}
  &M =_{\omega} N}
 &\mathbf{I} N \sim_{w\beta\Omega} N}
\]

\item Assume that the proof ends with an instance of the Leibnitz
Rule of the form:
\[\infer{[Q/z] X =_{\omega}[Q/z] Y}
        {[P/z] X =_{\omega} [P/z] Y\qquad P =_{\omega} Q}
\]
with $M \equiv [Q/z]X$ and $N \equiv [Q/z]Y$. By induction
hypothesis and Lemmata \ref{sym_trans} and \ref{context}, there are
proofs in normal form
of:
\[[P/z] X =_{\omega} [P/z] Y\quad,\quad
  [Q/z] X =_{\omega} [P/z] X\quad\hbox{and}\quad
  [P/z] Y =_{\omega} [Q/z] Y
\]
and therefore by Lemma \ref{sym_trans} again, we get a proof in
normal form of $[Q/z] X =_{\omega}[Q/z] Y$.\qed
\end{enumerate}

\noindent{\bf Remark.} Observe that the transformation of an instance
of the $\omega$-rule into a proof with an endpiece in normal form has
only an auxiliary character.  In other words, this transformation can
be done if needed, but {\em we do not want to perform it
systematically}.\medskip

\noindent{\it Proof of Theorem \ref{end-p-nf}.} It is clear that,
from the previous lemmas, Theorem \ref{end-p-nf} follows.\qed

\noindent {\bf Remark.} Theorem \ref{end-p-nf} is essentially a
particular case of a general result about Leibnitz Rule due to the
second author of the present paper. For more details the reader
should consult \cite{St1}.\medskip

Now consider a proof with an endpiece in normal form:
\[\infer{M =_{\omega} N}
  {\infer{M =_{\omega} M_t Q_t}
   {\infer{\dots\strut}
    {\infer{M =_{\omega} M_2 Q_2}
     {\infer{M =_{\omega} M_1 Q_1}
      {\infer{M =_{\omega} M_1 Q_1}
       {M =_{\omega} M
      &M \sim_{w\beta\Omega} M_1 P_1}
     &\mkern-16 mu P_1 =_{\omega} Q_1}
    &\mkern-16 mu M_1 Q_1 \sim_{w\beta\Omega} M_2 P_2}
   &\mkern-16 mu P_2 =_{\omega} Q_2}
  &\mkern-16 mu\hbox to 10 pt{\hfill}}
 &\mkern-16 mu M_t Q_t \sim_{w\beta\Omega} N}
\]
We represent this proof as a computation viz
\begin{equation}\label{star} M
\sim_{w\beta\Omega} M_1 P_1 =_{\omega} M_1 Q_1 \sim_{w\beta\Omega}
 M_2 P_2 ... \sim_{w\beta\Omega} M_t P_t =_{\omega} M_t Q_t
 \sim_{w\beta\Omega} N
\end{equation}

\noindent{\bf Remark.} We include the degenerate cases as follows:
\begin{enumerate}[$\bullet$]
\item $M =_{\omega} M$ gives rise to the computation $M
\sim_{w\beta\Omega} M$;
\item $M\sim_{w\beta\Omega} N$ can be directly considered as a computation.
\end{enumerate}

\begin{defi} We shall call the sequence (\ref{star}) {\em the end piece
computation} of a proof.
\end{defi}

\section{Ordinals}\label{ordinals}

 Since proofs are infinite
trees $\mathcal{T}$ they can be described by countable ordinals. In
the following, we shall need a few facts about countable ordinals.
For completeness, we recall the main notions involved.
For more details, see e.g. \cite{Sch}.

\medskip\noindent{\bf (a) Cantor Normal Form to the Base Omega} $(\omega)$.\
Every countable ordinal $\alpha$ can be written uniquely in the form
                    $\omega^{\alpha_1}*n_1 + ... + \omega^{\alpha_k}*n_k$
where $n_1 , ... , n_k$ are positive integers and $\alpha_1 > ... >
\alpha_k$ are ordinals.

Note that in the special case when $\alpha$ is a fixed point of
ordinal exponentiation (like $\epsilon_0$) we have
                     $\omega^\alpha*1$ as Cantor normal form of $\alpha$.

\medskip\noindent{\bf (b) Hessenberg Sum}.\
 Write $\alpha =  \omega^{\alpha_1}*n_1 + ...
+ \omega^{\alpha_k}*n_k$ and
       $\gamma =  \omega^{\alpha_1}*m_1 + ... + \omega^{\alpha_k}*m_k$
where some of the $n_i$ and $m_j$ may be 0. Then the {\em Hessenberg
Sum} is defined as follows: $ \alpha \oplus \gamma =_{def}
           \omega^{\alpha_1}*(n_1+m_1) + ... + \omega^{\alpha_k}*(n_k + m_k
           )$.

The Hessenberg sum is strictly increasing on both arguments. That is,
for $\alpha, \gamma$ different
from $0$, we have: $\alpha, \gamma < \alpha \oplus \gamma$.

\medskip\noindent{\bf (c) Hessenberg Product}\ .
            We only need this for product with an integer. We put:
            $\alpha \odot n =_{def} \alpha \oplus ... \oplus \  \alpha$
            $n$-times.
\medskip

 Coming back to proofs, observe first that we can assume that if a
proof has an endpiece, then this endpiece is in normal form (see the
previous Section). The ordinal that we want to assign to a proof
$\mathcal{T}$ (considered as a tree) is the transfinite ordinal
$ord(\mathcal{T})$, {\em the order of } $\mathcal{T}$, defined
recursively by:

\begin{defi}\label{ord}
 Let $\oplus$ be the Hessenberg  sum of
 ordinals defined above.
\begin{enumerate}[$\bullet$]
\item If
$\ \mathcal{T}$ ends in an endpiece computation of the form
(\ref{star}) and we are in the degenerate case $t= 0$ then
$ord(\mathcal{T}) =_{def} 1$;
\item If $\ \mathcal{T}$ ends in an instance of
the $\omega$-rule whose premisses have trees resp. $\mathcal{T}_{1},
\ldots \mathcal{T}_{i}, \ldots$ then $ord(\mathcal{T}) =_{def}
\omega ^{\theta}$, with $\theta = Sup\{ord(\mathcal{T}_1) \oplus ...
\oplus ord(\mathcal{T}_i) : i=1,2,...\}$;
\item If
$\ \mathcal{T}$ ends in an endpiece computation of the form
(\ref{star}), with $t > 0$,  and the equations $P_1 =_{\omega}
Q_1$,..., $P_t =_{\omega}Q_t$,
 have resp. trees $\mathcal{T}_{1}, \ldots, \mathcal{T}_{t}$
then $ord(\mathcal{T}) =_{def} 1 \oplus ord(\mathcal{T}_1) \oplus
... \oplus ord(\mathcal{T}_t) $.
\end{enumerate}
\end{defi}

\begin{fact}
If $\ \mathcal{T}$ ends in an endpiece computation of the form
(\ref{star}), with $t > 0$, and the equations $P_1 =_{\omega}
Q_1$,..., $P_t =_{\omega}Q_t$,
 have resp. trees $\mathcal{T}_{1}, \ldots, \mathcal{T}_{t}$
 then $ord(\mathcal{T})
> ord(\mathcal{T}_i)$, for each $i = 1, ..., t $.
\end{fact}

\proof $ord(\mathcal{T}_i) > 0$ and $\oplus$ is strictly
increasing on its arguments.\qed

\begin{fact}
Assume that $\mathcal{T}$ ends in an instance of the $\omega$-rule
whose premisses have, respectively, trees  $\mathcal{T}_1,\dots,
\mathcal{T}_t$,
... Then for any integers $t,n_1,\dots ,n_t $
\[ord(\mathcal{T}) > ord(\mathcal{T}_1)\odot n_1  \oplus ...
 \oplus  ord(\mathcal{T}_t)\odot n_t\ .\]
\end{fact}

\proof Let $ord(\mathcal{T}_i) = \alpha_i$, for $1 \leq i \leq
t$ and put all $\alpha_1, ... , \alpha_t$
into Cantor normal form:
\[\alpha_1 = \omega^{\beta_ 1} * n_{11} + ... + \omega^{\beta_ k} * n_{1k}
\quad\dots\quad
\alpha_t = \omega^{\beta_ 1} * n_{t1} + ... + \omega^{\beta_ k} *
n_{tk}\ .\]
Let $n = max \{ n_r, n_{j1} \} + 1$, with $j,r = 1 ... t$ . Then
\[\alpha_1 \odot n_1  \oplus ... \oplus  \alpha_t \odot n_t
 <\alpha_1 \odot n    \oplus ... \oplus  \alpha_t \odot n
 =\alpha_1 \oplus ... \oplus \alpha_t) \odot n \leq \omega^{\beta_1}
  *n^{2}*t\ .\]
Now let $\theta = Sup\{ord(\mathcal{T}_1) \oplus ... \oplus
ord(\mathcal{T}_i) : i=1,2,...\}$.
 We have $\omega^{\beta_1} <
\theta \leq \omega^{\theta} = ord(\mathcal{T})$.
But $ord(\mathcal{T})$ is a countable ordinal of the form
$\omega^\gamma$ and is thus closed under addition. Hence
$\omega^{\beta_ 1} *n^{2}*t < ord(\mathcal{T})$.\qed


\section{Cascades of Beta Reductions}

Recall that, as usual, we consider only closed terms. We define the
set of {\em weak $\beta\mathbf{\Omega}$ head normal forms (whnf)} as
follows:
\begin{enumerate}[(1)]
\item an unsolvable term is in whnf iff it is $\mathbf{\Omega}$;
\item
a solvable term is in whnf iff it has not a head weak $\beta$-redex
that is it has not the form
 $\lambda x_1 \ldots x_n. \: (\lambda x. U)V M_1 \cdots M_k$, with $(\lambda x. U)$
 and $V$ closed.
\end{enumerate}

\noindent Now we want to prove that the set weak $\beta\mathbf{\Omega}$ head
normal forms is {\em cofinal} w.r.t. weak
$\beta\mathbf{\Omega}$-reduction, in the sense of the following
theorem.

\begin{thm}For every $M$ there exists an $N$ in whnf, such that
$ M \longrightarrow^{*}_{w\beta\Omega} N$
\end{thm}

\proof
If $M$ is unsolvable, then $ M \longrightarrow^{*}_{w\beta\Omega}
\mathbf{\Omega}$.

Assume $M$ solvable.  Then $ M \longrightarrow^{*}_{\beta} M'$ by a
sequence of head $\beta$-reductions, where $M'$ has the form
$\lambda x_1 \ldots x_n. x_i V_1 \ldots V_m$ (see \cite{Ba84}
8.3.11). If every $\beta$-reduction is a weak one then take $N
\equiv M'$, otherwise $N$ is the first term in the sequence where a
weak head $\beta$-reduction cannot be performed.\qed

 By the Church-Rosser theorem for weak
$\beta\mathbf{\Omega}$-reductions, an endpiece computation
\begin{equation} M
\sim_{w\beta\Omega} M_1 P_1 =_{\omega} M_1 Q_1 \sim_{w\beta\Omega}
 M_2 P_2 ... \sim_{w\beta\Omega} M_t P_t =_{\omega} M_t Q_t
 \sim_{w\beta\Omega} N
\end{equation}
can be put in the form (that we still call an {\em endpiece
computation})
\[M \longrightarrow^{*}_{w\beta\Omega} \:\: R_{1} \:\:\:
 ^{*}_{w\beta\Omega}\!\longleftarrow M_1 P_1 =_{\omega} M_1 Q_{1}
\longrightarrow^{*}_{w\beta\Omega} \:\: R_{2} \:\:\:
^{*}_{w\beta\Omega}\!\longleftarrow M_2 P_2 \]
\[ ... \longrightarrow^{*}_{w\beta\Omega} M_t P_t =_{\omega} M_t Q_t
 \longrightarrow^{*}_{w\beta\Omega} \:\: R_{t+1} \:\:\: ^{*}_{w\beta\Omega}\!\longleftarrow N\]
Now, we want to show that special conditions can be imposed on the
weak  $\beta\mathbf{\Omega}$-reductions occurring in each endpiece
computation.

\begin{defi}\label{casc_end_piece}

An endpiece computation of the form
\[M \longrightarrow^{*}_{w\beta\Omega} \:\: R_{1} \:\:\:
 ^{*}_{w\beta\Omega}\!\longleftarrow M_1 P_1 =_{\omega} M_1 Q_{1}
\longrightarrow^{*}_{w\beta\Omega} \:\: R_{2} \:\:\:
^{*}_{w\beta\Omega}\!\longleftarrow M_2 P_2\]
\[ ... \longrightarrow^{*}_{w\beta\Omega} M_t P_t =_{\omega} M_t Q_t
 \longrightarrow^{*}_{w\beta\Omega} \:\: R_{t+1} \:\:\: ^{*}_{w\beta\Omega}\!\longleftarrow N\]
is called a {\em a cascade of weak $\beta\mathbf{\Omega}$
-reductions} iff
\begin{enumerate}[(1)]
\item all the confluence terms $R_{i}$, $1 \leq i \leq t+1$ are in
$whnf$;
 \item all the reductions of the form $R_{i} \:\:\:
 ^{*}_{w\beta\Omega}\!\longleftarrow M_i P_i$, with $1 \leq i \leq t$
   occurring in the
 endpiece are
 one step $\beta$-reductions of the form
 $[P_{i}/x]X \:\:\: _{w\beta\Omega}\longleftarrow (\lambda x. X)P_{i}$,
for some $X$, and moreover such $X$ has not the form $\lambda y_1
\ldots y_r. x X_1 \cdots X_m$.
 \end{enumerate}
\end{defi}

\noindent Note that this puts no restriction on left facing arrows.\\

\begin{defi}\label{casc}

The notion of a {\em cascaded proof} is defined inductively as
follows.
\begin{enumerate}[(1)]
 \item A proof with a degenerate endpiece is a {\em cascaded proof} if
 it has the form:
\[M \longrightarrow^{*}_{w\beta\Omega} \:\: R \:\:\:
 ^{*}_{w\beta\Omega}\!\longleftarrow N\]
 with $R$ in whnf.
 \item A proof ending
with an instance of the $\omega$-rule is a {\em cascaded proof} if
the proofs of the premisses of the instance are cascaded.
\item
Otherwise a proof is  {\em cascaded} if its endpiece is a cascade of
weak $\beta\mathbf{\Omega}$-reductions and all the proofs of the
leaves which are direct conclusions of the $\omega$-rule are
cascaded.
 \end{enumerate}
\end{defi}

In the following, we need the following well known fact about
$\mathcal{H}\mathbf{\omega}$.

\begin{prop}\label{b_tree}
If $M =_{\omega} N$ then $\mathrm{BT(}M\mathrm{)}=_{\eta}
\mathrm{BT(}N\mathrm{)}$, that is $M$ and $N$ have $\eta$-equal
B\"{o}hm trees.
\end{prop}

\proof
By Proposition 16.2.7 of \cite{Ba84}, this holds for equality in the
theory $\mathcal{H}^{*}$. Moreover, by Section 17.2 of \cite{Ba84},
we have that $\mathcal{H}\mathbf{\omega}$ is included in
$\mathcal{H}^{*}$.\qed

Now, we want to prove the following important fact about cascaded
proofs.

\begin{prop}\label{P:cascaded}
 If $\ M =_{\omega} N$ then there is a cascaded proof of $\ M =_{\omega} N$.
\end{prop}

\proof We prove this proposition by induction on the ordinal
$ord(\mathcal{T})$ of a proof $\mathcal{T}$ in normal form of $M
=_{\omega} N$.

For the base case just suppose that $M \sim_{w\beta\Omega} N$ and use
the Church-Rosser theorem.

\medskip\noindent{\bf Induction step.}  Assume first that $M =_{\omega} N$ is
the direct conclusion of the $\omega$-rule. This follows directly from
the induction hypothesis.

Otherwise, $M =_{\omega} N$ is the conclusion of a chain of equality
inferences:
\[M \sim_{w\beta\Omega} M_1 P_1 =_{\omega} M_1 Q_1 \sim_{w\beta\Omega} M_2 P_2
         \sim_{w\beta\Omega} ...
 \sim_{w\beta\Omega}
          M_t P_t =_{\omega} M_t Q_t \sim_{w\beta\Omega} N\]
where $t > 0$ and each $M_i P_i =_{\omega} M_i Q_i$ is the conclusion
of an instance of the $\omega$-rule. Again by the Church-Rosser
theorem we have the following computation:
\[M \longrightarrow^{*}_{w\beta\Omega} R_1 \:\:\:
 ^{*}_{w\beta\Omega}\!\longleftarrow  M_1 P_1 =_{\omega} M_1 Q_1
 \longrightarrow^{*}_{w\beta\Omega} R_2 \:\:\:
 ^{*}_{w\beta\Omega}\!\longleftarrow  M_2 P_2 \longrightarrow^{*}_{w\beta\Omega}\]
  \[... \longrightarrow^{*}_{w\beta\Omega} R_{t+1} \:\:\:
 ^{*}_{w\beta\Omega}\!\longleftarrow N\ .\]
Clearly each $R_i$ can be replaced by any weak
$\beta\mathbf{\Omega}$-reduct of $R_i$.

 Consider a reduction from $M_1 P_1$ to $R_1$ with all the weak
$\mathbf{\Omega}$-reductions (that is reductions of the form $
\longrightarrow_{\Omega}$) at the end; such a reduction exists by
\cite{Ba84} Proposition 15.2.9. Moreover, we can assume that no term
in the reduction is unsolvable, for otherwise $M$ and $N$ are both
unsolvable, by the previous proposition, and we simply have the
cascaded proof $M \longrightarrow_{w\beta\Omega} \mathbf{\Omega}
\:_{w\beta\Omega}\!\longleftarrow N$.

  We follow
all etraces of $P_1$ in the reduction of $M_1 P_1$ to $R_1$
attempting to simulate this with a reduction of $M_1 Q_1$. On the
$M_1 Q_1$ side we skip reductions internal to etraces of $P_1$. When
we come to redexes $(\lambda u.U) V$ where $P_1
\longrightarrow^{*}_{w\beta\Omega} \lambda u.U$, let $V = [V_{1} /
x_1 , ... , V_r / x_r ]X$ showing all the etraces of $P_1$ in $V$.
Then:

\medskip
\centerline{(*)\hfill $ Q_1 ([Q_1 / x_1 , ... , Q_1 / x_r ]X) =_{\omega}[([Q_1 /x_1
, ... , Q_1 /x_r ]X)/u ]U$\hfill}
\medskip
\noindent ({\em via} the equality $Q_1 ([Q_1 / x_1 , ... , Q_1 / x_r
]X) =_{\omega} P_1 ([Q_1 / x_1 , ... , Q_1 / x_r ]X)$) by a proof with
ordinal (much) less than $ord(\mathcal{T})$.  So, in the $M_1 Q_1$
side we replace the reduction of $(\lambda u.U) V$, taking place in
the $M_1 P_1$ side, with the computation:
\[Q_1 ([Q_1 / x_1 , ... , Q_1 / x_r ]X) =_{\omega} P_1 ([Q_1 / x_1 ,
... , Q_1 / x_r ]X) \longrightarrow_{w\beta\Omega} [[Q_1 / x_1 , ...
, Q_1 / x_r ]X / u]U\ .\]

Assume now that, in the $M_1 P_1$ side, we come to an
$\mathbf{\Omega}$-reduction containing etraces of $P_1$, say of the
form $U \longrightarrow_{w\beta\Omega} \mathbf{\Omega}$. Then if we
replace every occurrence of etraces of $P_1$ in $U$ with $Q_1$, we
obtain - by the previous proposition - a term $U'$ which is also
unsolvable. So, in the $M_1 Q_1$ side, we perform the reduction $U'
\longrightarrow_{w\beta\Omega} \mathbf{\Omega}$.

In the end we obtain $R_1$ as $[V_{1} / x_1 , ... , V_r / x_r ]X$
for some $X$ where $V_1 , ... , V_r$ are the remaining etraces of
$P_1$. On the $M_1 Q_1$ side we obtain
               $[Q_1 / x_1 , ... , Q_1 / x_r ]X$.
Since there are only finitely many instances of (*), we have that
             $[Q_1 / x_1 , ... , Q_1 / x_r ]X =_{\omega} N$
by a proof with ordinal $< ord(\mathcal{T})$ (use Fact 2 of the
Section \ref{ordinals}). Thus there exists a cascaded proof
$\mathcal{T}^{+}$ of
                 $[Q_1 / x_1 , ... , Q_1 / x_r ]X =_{\omega} N$.

{\em Subcase 1.}\ $X \equiv x$ so that $R_1$ is an etrace of $P_1$.\\
Let $L$ be given. Since there are only finitely many instances of
(*), we have that $ML =_{\omega} NL$ by a proof with ordinal $<
ord(\mathcal{T})$ (again, use Fact 2 of the Section \ref{ordinals}).
Thus there exists a cascaded proof of $ML =_{\omega} NL$. Since this
holds for every $L$ we obtain a cascaded proof $M =_{\omega} N$, by
an application of the $\omega$-rule with cascaded proofs for all the
premisses.

{\em Subcase 2.}\ Otherwise.\\
By the Church-Rosser theorem there exists a common reduct $V$ of all
the $V_1 , ... , V_r$. In addition, by induction hypothesis, there
exists a cascaded proof $\mathcal{T}^{++}$ of $V =_{\omega} Q_1$. We
distinguish two cases.

{\em Subcase 2.1.}\ $X$ begins with some variable $x_j$, say $X \equiv
x_j X_1 \ldots X_s$.\\
In this case, since there are only finitely many instances of (*),
 to which we add a proof of:
\[V_j ([Q_{1} / x_1 , ... , Q_{1} / x_r ]X_1)... ([Q_{1} / x_1 , ...
, Q_{1} / x_r ]X_s) =_{\omega}\]
\[=_{\omega} Q_{1}([Q_{1} / x_1 , ... , Q_{1} / x_r ]X_1)... ([Q_{1}
/ x_1 , ... , Q_{1} / x_r ]X_s)\]
 we have that
$V_j ([Q_{1} / x_1 , ... , Q_{1} / x_r ]X_1)... ([Q_{1} / x_1 , ...
, Q_{1} / x_r ]X_s) =_{\omega} N$ has a proof with ordinal $<
ord(\mathcal{T})$. So, there exists a cascaded proof
$\mathcal{T}^{*}$ of this equality.

Thus, in this case, the desired cascaded proof of $M =_{\omega} N$
is obtained concatenating the following pieces:
\begin{enumerate}[(1)]
\item $M \longrightarrow^{*}_{w\beta\Omega}
V_j ([V / x_1 , ... , V / x_r ]X_1)... ([V / x_1 , ... , V / x_r
]X_s)$
\item
$V_j ([V / x_1 , ... , V / x_r ]X_1)... ([V / x_1 , ... , V / x_r
]X_s)\:\:\:_{w\beta\Omega}\!\longleftarrow$\\
$ \:\:\:_{w\beta\Omega}\!\longleftarrow (\lambda x.V_j ([x / x_1 ,
... , x / x_r ]X_1)... ([x / x_1 , ... , x / x_r ]X_s))V$
\item
$(\lambda x.V_j ([x / x_1 , ... , x / x_r ]X_1)... ([x / x_1 , ... ,
x / x_r ]X_s))V =_{\omega}$\\
$(\lambda x.V_j ([x / x_1 , ... , x / x_r ]X_1)... ([x / x_1 , ... ,
x / x_r ]X_s))Q_1$
\item $(\lambda x.V_j ([x / x_1 , ... , x / x_r ]X_1)... ([x / x_1 , ... ,
x / x_r ]X_s))Q_1 \longrightarrow_{w\beta\Omega}$\\
$\longrightarrow_{w\beta\Omega} V_j ([Q_{1} / x_1 , ... , Q_{1} /
x_r ]X_1)... ([Q_{1} / x_1 , ... , Q_{1} / x_r ]X_s)$
\item $V_j ([Q_{1} / x_1 , ... , Q_{1} / x_r ]X_1)...
([Q_{1} / x_1 , ... , Q_{1} / x_r ]X_s) =_{\omega} N$
   \end{enumerate}
Observe that $V_j ([V / x_1 , ... , V / x_r ]X_1)... ([V / x_1 , ...
, V / x_r ]X_s)$ is still in whnf since this class is closed under
internal reductions. This ends the proof of Subcase 2.1.

 {\em Subcase 2.2.}\ Otherwise.\\
Then the endpiece of the desired cascaded proof is as follows:
\begin{enumerate}[(1)]
\item $M \longrightarrow^{*}_{w\beta\Omega}
[V / x_1 , ... , V / x_r ]X$
\item
$[V / x_1 , ... , V / x_r ]X \:\:\:_{w\beta\Omega}\!\longleftarrow$\\
$ \:\:\:_{w\beta\Omega}\!\longleftarrow (\lambda x.[x / x_1 , ... ,
x / x_r ]X) V$
\item
$(\lambda x.[x / x_1 , ... ,
x / x_r ]X) V =_{\omega}$\\
$(\lambda x.[x / x_1 , ... , x / x_r ]X) Q_1$
\item $(\lambda x.[x / x_1 , ... , x / x_r ]X) Q_1 \longrightarrow_{w\beta\Omega}$\\
$\longrightarrow_{w\beta\Omega} [Q_{1} / x_1 , ... , Q_{1} / x_r ]X$
\item $[Q_{1} / x_1 , ... , Q_{1} / x_r ]X  =_{\omega} N$
\end{enumerate}
This ends the proof of Subcase 2.2, and the proof of Proposition
\ref{P:cascaded} is complete.\qed

In the following lemma, we recall that $M$ and $N$ (possibly with
indexes) always stand for closed terms.

\begin{lem}\label{open} Suppose that:
\begin{enumerate}[\em(1)]
\item $U_1 , U_2$
 contain
the free variable $u$ and no other free variable;
\item $V_1 , V_2$
 contain
at most the free variable $u$ and no other free variable;
\item $V_1 U_1 M_1 ... M_m$ and $V_2U_2 N_1
... N_m$ are solvable;
\item $\mathcal{T}$ is  a cascaded proof, not
ending in the $\omega$-rule, of
                  $\lambda u. V_1 U_1 M_1 ... M_m =_{\omega} \lambda u. V_2 U_2 N_1 ... N_m$.
\end{enumerate}
       Then for each $i$, with $1 \leq i \leq m$,
                        $M_i =_{\omega} N_i$.
\end{lem}

\proof By induction on $ord(\mathcal{T})$.

\medskip\noindent{\bf Base case.}
  $ord(\mathcal{T}) = 1$.  In this case no head
  $\beta$-redex with a reduct of $U_i$, $i = 1,2$, as the argument can
  be contracted as a weak $\beta$-redex. Neither $U_i$ can be part of
  a head weak $\mathbf{\Omega}$-redex. Thus the proof contains weak
  $\beta\mathbf{\Omega}$-conversions of the $M_i$ to the $N_i$.

\medskip\noindent{\bf Induction step.}
  $ord(\mathcal{T})$ is infinite.
          We can freely assume that $\mathcal{T}$ has the form:
\[\lambda u. V_1 U_1 M_1 ... M_m  \longrightarrow^{*}_{w\beta\Omega}
 R_1
 \:\:\:_{w\beta\Omega}\!\longleftarrow  L_1 P_1 =_{\omega}\]
\[=_{\omega} L_1 Q_1
 \longrightarrow^{*}_{w\beta\Omega} R_2 \:\:\:_{w\beta\Omega}\!\longleftarrow
   L_2 P_2 =_{\omega} L_2 Q_2
 \longrightarrow^{*}_{w\beta\Omega} \ ... \ L_t P_t =_{\omega}\]
\[=_{\omega} L_t Q_t
  \longrightarrow^{*}_{w\beta\Omega} R_{t+1}
\:\:\:_{w\beta\Omega}^{*}\!\longleftarrow \lambda u. V_2U_2N_1 ...
N_m\ .\]
We claim that each $R_j$, with $1 \leq j \leq t+1$, must have the
form $\lambda u. V'_{1j} U'_{1j} M'_{1j} ... M'_{mj}$, for some
$V'_{1j} , U'_{1j} , M'_{1j} , ..., M'_{mj}$, with $\lambda u.
V'_{1j} =_{\omega} \lambda u. V_1$, $\lambda u. U'_{1j} =_{\omega}
\lambda u. U_1$, and, for
$1 \leq i \leq m$, $M'_{ij} =_{\omega} M_i$.

To prove the claim observe that it is true for $R_1$.
 So let $R_1 \equiv \lambda u. V'_{11} U'_{11} M'_{11} ... M'_{m1}$. Consider
now $R_2$. Since $\mathcal{T}$ is cascaded, the reduction $R_1
 \:\:\:_{w\beta\Omega}\!\longleftarrow  L_1 P_1$ implies that
$L_1$ has the form $\lambda x u. V^{*}_1 U^{*}_1 X^{*}_1 ...
X^{*}_m$ and that $V'_{11} \equiv [P_1 / x]V^{*}_1$, $U'_{11} \equiv
[P_1 / x]U^{*}_1$, and, for $1 \leq i \leq m$, $M'_{i1} \equiv [P_1
/ x]X^{*}_i$. So, it is clear that $R_2$ has the required form.
Repeating this argument we get the claim.

By the claim, it follows that $\mathcal{T}$ contains a proof of $M_i
=_{\omega} N_i$ for each $i$, with $1 \leq i \leq m$.\qed

\section{Barendregt's Construction}\label{BC}

The present Section requires acquaintance with Section 17.4 of
\cite{Ba84}. However, we will modify Barendregt's construction in a
number of minor points, in order to have a better control of the
behavior of the terms. On the other hand, the two constructions are
almost identical, and we hope that the reader could be able to
reconstruct the correspondences between them.\\ Assume that an
effective coding of finite sequences of natural numbers with natural
numbers has been fixed. We call the coding numbers {\em sequence
numbers} and we denote them by symbols $s$, $s'$, etc. We write $s'
\leq s$ ($s' < s$) to denote that $s'$ is a subsequence (resp. a
proper subsequence) of $s$. Let $f$ be a function from natural numbers
to natural numbers; following again \cite{Ba84}, we denote by
$\bar{f}(n)$ the sequence number of the sequence $\langle f(0), \ldots
, f(n-1)\rangle$. Now, let $P(n)$ be a $\mathbf{\Pi_{1}^{1}}$
predicate. Then: \[ P(n) \Longleftrightarrow \forall f \: \exists m \:
R(\bar{f}(m), n)\] for some recursive relation $R$. A sequence number
$s$ is {\em $n$-secured} iff $\exists s' < s. \: R(s', n)$, otherwise
{\em $n$-unsecured}. Observe that, for $n$ fixed the set of
$n$-unsecured sequence numbers is closed under the subsequence
relation and therefore is a {\em tree} (possibly empty). Thus $P(n)$
holds iff this tree is {\em well-founded}, i.e. not $s_0 < s_1 < s_2
\cdots $ for some infinite sequence of $n$-unsecured sequence numbers.
Moreover, the notion "$s$ is $n$-unsecured" is recursive. We can sum
up our discussion by the following well known theorem (see
\cite{Rogers} Ch.16 Th.20).

\begin{thm}
The set of (indices of) well founded recursive trees is
$\mathbf{\Pi_{1}^{1}}$-complete.\qed
\end{thm}

Now, let again the $\mathbf{\Pi_{1}^{1}}$ predicate $P(n)$ be fixed.
Let $n$ be fixed once for all, we denote by $T$ the tree of all
$n$-unsecured sequence numbers. Now we recall (a version of) Lemma
17.4.11 of \cite{Ba84}.

\begin{lem}\label{box}
There is a closed term $\Box$ such that:
\[ \Box\underline{s} =_{\omega} \left \{ \begin{array}{ll}
\mathbf{K^*} & \mbox{if $s \in T$}\\
\mathbf{\Omega} &  \mbox{otherwise}
\end{array} \right. \]
where $\mathbf{K^*} \equiv \lambda ab. b$.
\end{lem}

\proof
The lemma follows from the fact that $T$ is recursive (see 16.1.10
of \cite{Ba84}).\qed

As shown by Lemma 17.4.11 of \cite{Ba84}, one can have a term $\Box$
which is uniform in $n$, i.e. such that, given $n$, it returns a term
representing the corresponding tree. More in general, Barendregt shows
that all the construction can be done uniformly in $n$. To simplify a
little the construction, we have everywhere suppressed this
dependency.  This will not affect our results.

On the other hand, we need the following slightly stronger version
of the previous lemma:

\begin{lem}
There is a closed term $D$ such that:
\begin{enumerate}[\em(1)]
\item
For every numeral $\underline{m}$, $D x \underline{m}$ has a $\beta
\mathbf{\Omega}$-normal form beginning with $x$ and containing
$\underline{m}$. (Where $\underline{m}$ is a parameter needed in the
following).
\item For every $\underline{s}$
                        \[ D\underline{s}  =_{\omega} \left \{ \begin{array}{ll}
\mathbf{K^*} & \mbox{if $s \in T$}\\
\mathbf{\Omega} &  \mbox{otherwise}
\end{array} \right.\]
\item if $s$ belongs to the tree, then $D\underline{s}
\longrightarrow^{*}_{w\beta\Omega}
\mathbf{K^*}$ by head weak $\beta$-reductions.
\end{enumerate}
\end{lem}

\proof
First of all, we can assume that $\Box$ of Lemma \ref{box}, has the
property that if $s$ belongs to the tree, then $\Box\underline{s}
\longrightarrow^{*}_{w\beta\Omega} \mathbf{K^*}$ by head weak
$\beta$-reductions. This can be obtained by the representation of
recursive functions by $\lambda$-terms. We can also assume that
$\Box$ has the form $\lambda x. X$. Transform the term $X$ as in
\cite{St2}, by replacing (inside out) each $\beta$-redex in $X$ of
the form $(\lambda z. Z)W$ into the term $x
\mathbf{I}\mathbf{I}(\lambda z. Z)W$, where $\mathbf{I}$ is the
identity combinator. Let $Y$ be the resulting term, which is
obviously in $\beta$-normal form and let $D \equiv \lambda x. x
\mathbf{I}\mathbf{I}Y$. Observe that for every term $\underline{s}$
(actually a numeral) representing a sequence
number we have that:
\[\underline{s}\mathbf{I}\mathbf{I} \longrightarrow^{*}_{w\beta\Omega}
 \mathbf{I}\ ,\quad\hbox{by head weak $\beta$-reductions,}\]
so that for every $\underline{s}$:
\[D\underline{s} \longrightarrow^{*}_{w\beta\Omega} \mathbf{K^*}\ ,
\quad\hbox{by head weak $\beta$-reductions, if $s \in T$,}\]
  and:
\[D\underline{s} \longrightarrow^{*}_{w\beta\Omega}
\mathbf{\Omega}\ ,\quad\hbox{otherwise.}\]
Finally, it is obvious that for every numeral $\underline{m}$, $D x
\underline{m}$ has a $\beta \mathbf{\Omega}$-normal form beginning
with $x$ and containing $\underline{m}$.\qed

We now come back to the representation of sequences. For simplicity
we denote the term representing the concatenation function by the
infixed operator $*$. We can freely require that $\underline{s}*z$
has a $\beta\mathbf{\Omega}$ normal form beginning with $z$
 (using the same technique of the previous proof).

Now, we define several terms.

\begin{enumerate}[(1)]
\item $\mathbf{\Theta} \equiv (\lambda ab. b(aab)) (\lambda ab.
b(aab))$ (Turing's fixed point).

\item $Z \equiv \mathbf{\Theta}(\lambda axf. [fx ,\ x(\lambda
  u.u\mathbf{\Omega})y(a(x^{+})f)])$, where by $x^{+}$ we denote the
  application of the {\em successor function} to $x$.  Observe that
  $y$ is free in $Z$.
\item $F_0 \equiv \lambda wxy. D x\underline{0}(\lambda ab. b(aab))(\lambda ab.
b(aab))(\lambda axf. [fx , \  x(\lambda
u.u\mathbf{\Omega})y(a(x^{+})f)])\underline{0}(\lambda z. w(x*z))$
\item $B_0 \equiv \mathbf{\Theta}F_0$
\item $F_1 \equiv \lambda wxy. D x\underline{1}(\lambda ab. b(aab))(\lambda ab.
b(aab))(\lambda axf. [fx , \  x(\lambda
u.u\mathbf{\Omega})y(a(x^{+})f)])\underline{0}(\lambda z. w(x*z))$
\item $B_1 \equiv \mathbf{\Theta}F_1$
\end{enumerate}

First of all observe that both $\mathbf{\Theta}$ and $Z$ are not
subterms of $F_0$ (and neither of $F_1$); however these terms are
generated during the reduction of $F_0$ and $F_1$ (see below). To
relate the previous definitions to \cite{Ba84} page 463, we first
observe that $Z$ (with suitable arguments) behaves like the term
$\Pi$ of 17.4.8 of \cite{Ba84}. Indeed the following lemma holds.
\begin{lem}\label{zeta}
For all $M$, $N$ and natural number $m$ the following are equivalent:
\begin{enumerate}[$\bullet$]
\item for every $P$, $([P/y]Z)\underline{m}M =_{\omega}
([P/y]Z)\underline{m}N$
\item for every natural number $m'$, with $
m \leq m'$ $M\underline{m'} =_{\omega} N\underline{m'}$.
\end{enumerate}
 \end{lem}

\proof
Notice that for every $P$ and $M$, and for every $m$:\\
$([P/y]Z)\underline{m}M =_{\omega} [M \underline{m},\:
P\mathbf{\Omega}^{\sim m}(([P/y]Z)\underline{m}^{+}M)]$\\
(where notations $\sim m$ and $[M_1 ,M_2, \ldots]$ are as in
\cite{Ba84} page 25, and, respectively, page 169) then argue as in
Theorem 17.4.9 of \cite{Ba84}.\qed

Now, we consider the behavior of $B_0$ and $B_1$, which correspond
(with minor modifications) to the terms $B^{n}_0$ and, respectively,
$B^{n}_1$ of 17.4.13 of \cite{Ba84}.

The terms $B_0$ and $B_1$ have the same behavior and are
distinguishable only by the passive parameters $\underline{0}$ and,
respectively, $\underline{1}$.

Consider, e.g., $B_0$. We have, with 3 head reduction steps:
\[\eqalign{
 B_0
&\longrightarrow_{w\beta\Omega}(\lambda b. b(\mathbf{\Theta}b))F_0\cr
&\longrightarrow_{w\beta\Omega} F_0(\mathbf{\Theta}F_0)\cr
&\longrightarrow_{w\beta\Omega} \lambda xy. D
x\underline{0}(\lambda ab. b(aab))(\lambda ab.b(aab))(\lambda axf.%
[fx , \ x(\lambda
u.u\mathbf{\Omega})y(a(x^{+})f)])\underline{0}(\lambda z. B_0(x*z))
}\] and if $s$ belongs to the tree then, with a sequence of head
weak $\beta$-reductions:
\[\eqalign{
 B_0 \underline{s}
&\longrightarrow^{*}_{w\beta\Omega} \lambda y. D
  \underline{s}\underline{0}(\lambda ab. b(aab))(\lambda ab.%
  b(aab))(\lambda axf. [fx , \ x(\lambda
  u.u\mathbf{\Omega})y(a(x^{+})f)])\underline{0}(\lambda z.%
  B_0(\underline{s}*z))\cr
&\longrightarrow^{*}_{w\beta\Omega}
  \lambda y. \mathbf{\Theta}(\lambda axf. [fx , \  x(\lambda
  u.u\mathbf{\Omega})y(a(x^{+})f)])\underline{0}
  (\lambda z. B_0(\underline{s}*z))\cr
&\longrightarrow^{*}_{\beta}\hbox{(head $\beta$-reductions)}
  \lambda y.[(\lambda z. B_0(\underline{s}*z))\underline{0} ,
  \underline{0}(\lambda u.u\mathbf{\Omega})(y(Z(\underline{0}^{+})
  (\lambda z. B_0(\underline{s}*z))))]\cr
&\longrightarrow^{*}_{w\beta\Omega} \lambda y.
  [B_0(\underline{s}*\underline{0}) , \
  y(Z(\underline{0}^{+})(\lambda z.B_0(\underline{s}*z)))]
}\]
We refer to the reduct
           $\lambda y.[(\lambda z. B_0(\underline{s}*z))\underline{0}, \
            \underline{0}(\lambda u.u\mathbf{\Omega})(y(Z(\underline{0}^{+})
            (\lambda z. B_0(\underline{s}*z))))]$
as the {\em pivot point}, and similarly for $B_1$ and
$B_1\underline{s}$. So, a head reduction of $B_0\underline{s}$ or
$B_1\underline{s}$ begins with 3 head reductions followed by a head
reduction of $D \underline{s}$ which either terminates in
$\mathbf{K^*}$ or fails to terminate. In the first case the next
head reduction is of the $\mathbf{K^*}\underline{i}$ ($i = 0,1$)
redex followed by $\mathbf{I}(\lambda ab. b(aab))
\longrightarrow_{w\beta\Omega} (\lambda ab. b(aab))$ and
$\mathbf{\Theta} \longrightarrow_{w\beta\Omega} \lambda b.
b(\mathbf{\Theta}b)$. In the weak $\beta\mathbf{\Omega}$ case this
is the end of the head reduction sequence since $y$ is contained in
the argument of the head redex. In unrestricted
$\beta\mathbf{\Omega}$-reduction there are 3 more reductions to the
pivot point. This ends the description of the behavior of terms
$B_0$ and $B_1$.

Let $T(s)$ denote the subtree of $T$ rooted at the sequence $s$.  Here
we include the empty tree, in case $s$ is not in $T$.  As in
\cite{Ba84} 17.4.14, we have that:

\begin{thm}\label{wf1} For every sequence $s$, if $T(s)$ is well founded
 then $B_0\underline{s} =_{\omega} B_1\underline{s}$
\end{thm}

\proof
Actually in \cite{Ba84} 17.4.14, this is proved for the empty
sequence $\langle\:\rangle$. However the same proof carries on,
since for every sequence $s'$, with $s \leq s'$ :
\begin{enumerate}[$\bullet$]
\item  if $s' \not\in T$
 then $B_0\underline{s'} =_{\omega} \mathbf{\Omega} =_{\omega}
 B_1\underline{s'}$ ;
\item  if $s' \in T$
 then, as shown above,
  $B_0\underline{s'} =_{\omega} \lambda y.[B_0(\underline{s'}*\underline{0}) , \
y(Z(\underline{0}^{+})(\lambda z.B_0(\underline{s'}*z)))]$ and
$B_1\underline{s'} =_{\omega}\lambda
y.[B_1(\underline{s'}*\underline{0}) , \
y(Z(\underline{0}^{+})(\lambda z.B_1(\underline{s'}*z)))]$.

On the other hand, by Lemma \ref{zeta}, if for every $m$,
$B_0(\underline{s'}
* \underline{m}) =_{\omega}
 B_1(\underline{s'} * \underline{m})$ then
for every $P$, $([P/y]Z)\underline{0}(\lambda
z.B_0(\underline{s'}*z)) =_{\omega} ([P/y]Z)\underline{0}(\lambda
z.B_1(\underline{s'}*z))$. By the $\omega$-rule it follows that:
$\lambda y. Z \underline{0}(\lambda z.B_0(\underline{s'}*z))
=_{\omega} \lambda y. Z \underline{0}(\lambda
z.B_1(\underline{s'}*z))$.

But $\lambda y. Z \underline{0}(\lambda z.B_0(\underline{s'}*z))
=_{\omega} B_0\underline{s'}$ and $\lambda y. Z
\underline{0}(\lambda z.B_1(\underline{s'}*z)) =_{\omega}
 B_1\underline{s'}$, and thus $B_0\underline{s'} =_{\omega}
 B_1\underline{s'}$.
\end{enumerate}
Now argue by bar induction as in \cite{Ba84} 17.4.14.\qed

\begin{thm}\label{wf2} For every sequence $s$, if
$B_0\underline{s} =_{\omega} B_1\underline{s}$ then $T(s)$ is well
founded.
\end{thm}

\proof By induction on the ordinal $ord(\mathcal{T})$ of a cascaded
proof $\mathcal{T}$ of $B_0\underline{s} =_{\omega}
B_1\underline{s}$. We shall assume that the weak head normal form
 restrictions on confluence terms are in effect.

\medskip\noindent{\bf Base case.} $ord(\mathcal{T}) =1$. Under the hypothesis
that $ord(\mathcal{T}) = 1$ we have that $B_0\underline{s}$ and
$B_1\underline{s}$ weak $\beta\mathbf{\Omega}$-convert. We shall
show that $s$ does not belong to $T$ and that $B_0\underline{s}
=_{\omega} \mathbf{\Omega} =_{\omega} B_1\underline{s}$. We proceed
by induction on the lengths of standard
$\beta\mathbf{\Omega}$-reductions to a common reduct (note here that
standardization does not in general hold for weak
$\beta\mathbf{\Omega}$-reduction so we revert to plain
$\beta\mathbf{\Omega}$). Assume that $s$ actually belongs to $T$.
First we show that both reductions must proceed all the way to the
pivot point. Clearly both head reductions must complete the head
reduction of $D \underline{s}$ and the reduction
$\mathbf{K^*}\underline{i} \longrightarrow_{w\beta\Omega}
\mathbf{I}$ to project the index $\underline{i}$ (where $i = 0,1$).
Since each succeeding term in the head reduction to the pivot point
has, respectively, 5, 4, 3, 2, 1 components, if both reductions to
the pivot point are not completed then they must stop to a term with
the same number of components. It follows that $\lambda
z.B_0(\underline{s}*z)$ and $\lambda z.B_1(\underline{s}*z)$ have
shorter standard $\beta\mathbf{\Omega}$-reductions to a common
reduct. But this is clearly impossible by the conditions on $D$ and
$*$, which imply that $\lambda z.B_0(\underline{s}*z)$ and $\lambda
z.B_1(\underline{s}*z)$ have B\"{o}hm trees which are not
$\eta$-equal. Thus both reductions proceed to the pivot point. Thus
there are shorter standard confluent $\beta\mathbf{\Omega}$-reductions from:
\[(\lambda z. B_0(\underline{s}*z))\underline{0}
  \quad\hbox{and}\quad
  (\lambda z.B_1(\underline{s}*z))\underline{0}\ ,
\]
and from:
\[\underline{0}(\lambda
  u.u\mathbf{\Omega})y(Z(\underline{0}^{+})(\lambda z.
  B_0(\underline{s}*z)))
  \quad\hbox{and}\quad
  \underline{0}(\lambda
  u.u\mathbf{\Omega})y(Z(\underline{0}^{+})(\lambda z.
  B_1(\underline{s}*z)))\ .
\]
 In particular by similar reasoning there exists shorter
confluent standard reductions from:
\[\mathbf{\Theta}(\lambda axf. [fx , \ x(\lambda
u.u\mathbf{\Omega})y(a(x^{+})f)])(\underline{0}^{+})(\lambda
z. B_0(\underline{s}*z))
\]
 and
\[\mathbf{\Theta}(\lambda axf. [fx , \
x(\lambda u.u\mathbf{\Omega})y(a(x^{+})f)])(\underline{0}^{+})(\lambda
z. B_1(\underline{s}*z))\ .
\]
Now we can repeat the above argument with minor modifications
forever since no $\mathbf{\Omega}$-reductions are possible. This is
impossible and proves that $s$ cannot be in $T$.

\medskip\noindent{\bf Induction step.}
 $ord(\mathcal{T})$ is infinite.

We distinguish twocases.

\medskip\noindent{\bf Case 1}. $\mathcal{T}$ ends with a direct
conclusion of the $\omega$-rule.

 Thus
for each closed term $M$ and any sequence $H_1 ... H_t$ of closed
terms:
\[B_0\underline{s} M H_1 ... H_t
  =_{\omega} B_1\underline{s} M H_1 ... H_t
\]
has a cascaded proof of ordinal smaller than $ord(\mathcal{T})$. To
see this observe that for any $M$, $B_0\underline{s} M
  =_{\omega} B_1\underline{s} M $
has a cascaded proof of ordinal, say $\gamma$, smaller than
$ord(\mathcal{T})$, which is - by definition of $ord(\mathcal{T})$ -
of the form $\omega^{\theta}$ for some $\theta > \gamma$. Actually,
for every $k$, $\theta > \gamma \oplus k$. Now, $B_0\underline{s} M
H_1 ... H_t
  =_{\omega} B_1\underline{s} M H_1 ... H_t$ can be obtained from
  the endpiece:
\[B_0\underline{s} M H_1 ... H_t \sim_{w\beta\Omega}
  (\lambda x.x H_1 ... H_t)(B_0\underline{s}M)
  =_{\omega} (\lambda x.x H_1 ... H_t)(B_1\underline{s}M)
  \sim_{w\beta\Omega} B_1\underline{s} M H_1 ... H_t\ ,
\]
which clearly has a cascaded proof of ordinal smaller than
$\gamma \oplus k$, for some $k$, and therefore smaller than
$\omega^{\theta}$ .

Now for any $m$, we can choose $M, H_1, ... H_{t_{m}}$ to
B\"{o}hm out, as in Theorem 17.4.9 of \cite{Ba84},
$B_0(\underline{s}*\underline{m})$ and
$B_1(\underline{s}*\underline{m})$ from $ B_0\underline{s}$
and, respectively, $ B_1\underline{s}$.

Thus for each $m$,
$B_0(\underline{s}*\underline{m}) =_{\omega}
B_1(\underline{s}*\underline{m})$ is provable  by a proof with
ordinal smaller than $ord(\mathcal{T})$. Hence, by induction
hypothesis, the subtree $T(s*m)$ of $T$ rooted at $s*m$ is well
founded. It follows that the subtree of $T$ rooted
at $s$ is well founded as well.

\medskip\noindent {\bf Case 2}. Otherwise.

 So $\mathcal{T}$ has an
endpiece.
Since $\mathcal{T}$ is cascaded, the endpiece has the form:
\[B_0\underline{s} \longrightarrow^{*}_{w\beta\Omega} R_1
\:\:\:_{w\beta\Omega}\!\longleftarrow  M_1 P_1 =_{\omega} M_1 Q_1
\longrightarrow^{*}_{w\beta\Omega} R_2
\:\:\:_{w\beta\Omega}\!\longleftarrow  M_2 P_2 =_{\omega} M_2 Q_2
\longrightarrow^{*}_{w\beta\Omega}\]
\[ ...
\longrightarrow^{*}_{w\beta\Omega} R_{t+1}
\:\:\: ^{*}_{w\beta\Omega}\!\longleftarrow B_1\underline{s}\ .
\]
Where each $R_i$ is in $whnf$ and all left-arrow reductions (with
the possible exception of the last one) are one step weak
$\beta$-reductions {\em not} of the form $(\lambda x y_1
\ldots y_r. x X_1 \cdots X_m)U$.

We shall show that also in this case $s$ is not in $T$. By
contradiction, assume $s$ in $T$. Since the weak head normal form
restrictions are in effect, the head reduction part of
the reduction to $R_1$ terminates in:
\[\lambda y. \lambda b.
b(\mathbf{\Theta}b)(\lambda axf. [fx , \  x(\lambda
u.u\mathbf{\Omega})y(af(x^{+}))])\underline{0}(\lambda
z. B_0(\underline{s}*z))
\]
 and similarly the head reduction part of the reduction from
$B_1\underline{s}$ to
$R_{t+1}$ terminates in:
\[\lambda y. \lambda b.
b(\mathbf{\Theta}b)(\lambda axf. [fx , \  x(\lambda
u.u\mathbf{\Omega})y(af(x^{+}))])\underline{0}(\lambda
z. B_1(\underline{s}*z))\ .
\]
Now, let:
\begin{enumerate}[$\bullet$]
\item $V_1 \equiv V_2 \equiv \lambda b. b(\mathbf{\Theta}b)$
\item $U_1 \equiv U_2 \equiv \lambda axf.
 [fx , \  x(\lambda u.u\mathbf{\Omega})y(af(x^{+}))]$
 \item $M_1 \equiv N_1 \equiv \underline{0}$
 \item $M_2 \equiv \lambda z. B_0(\underline{s}*z)$
\item $N_2 \equiv \lambda z. B_1(\underline{s}*z)$
\end{enumerate}

Thus by the Lemma \ref{open} there exists a proof of $\lambda z.
B_0(\underline{s}*z)=_{\omega} \lambda z. B_1(\underline{s}*z)$. But
this is impossible because these terms have B\"{o}hm trees which are
not $\eta$-equal. This completes the proof.\qed

\begin{cor} The set $\{ (M, N) |  M =_{\omega} N \}$ is
$\mathbf{\Pi_{1}^{1}}$-complete.
\end{cor}

\proof
Let $P(n)$ be a $\mathbf{\Pi_{1}^{1}}$ predicate. Given any natural
number $n$, to compute the truth value of $P(n)$, construct the
recursive tree $T$ of all $n$-unsecured sequence numbers. Then
construct the terms $B_0$ and $B_1$. Then use Theorem \ref{wf1} and
Theorem \ref{wf2} to determine ({\em via} equality in
$\mathcal{H}\mathbf{\omega}$) if $T$ is well founded.\qed

\section*{Acknowledgement}
We thank the anonymous referee for her/his help in substantially
improving a previous version of the paper.

\end{document}